\documentclass[pra,aps,twocolumn,amsmath,amssymb]{revtex4}

\usepackage{graphicx}
\usepackage{dcolumn}
\usepackage{bm}
\usepackage[extension=xxx]{hyperref}

\newcommand{\beq}{\begin{equation}}
\newcommand{\eeq}{\end{equation}}
\newcommand{\beqa}{\begin{eqnarray}}
\newcommand{\eeqa}{\end{eqnarray}}

\def\beq{\begin{equation}}

\usepackage{color}

\begin{document}

\title{Direct cooling in an optical lattice by amplitude modulation}
\date{\today}

\author{M. Arnal, V. Brunaud, G. Chatelain, C. Cabrera-Guti\'errez, E. Michon, P. Cheiney, J. Billy, D. Gu\'ery-Odelin}

\affiliation{Universit\'e de Toulouse ; UPS ; Laboratoire Collisions Agr\'egats R\'eactivit\'e, IRSAMC ; F-31062 Toulouse, France} 
\affiliation{CNRS ; UMR 5589 ; F-31062 Toulouse, France}

\begin{abstract}
We report on a generic cooling technique for atoms trapped in optical lattices. It consists in modulating the lattice depth with a proper frequency sweeping. This filtering technique removes the most energetic atoms, and provides with the onset of thermalization a cooling mechanism reminiscent of evaporative cooling. However, the selection is here performed in quasi-momentum space rather than in position space. Interband selection rules are used to protect the population with a zero quasi-momentum, namely the Bose Einstein condensate.  Direct condensation of thermal atoms in an optical lattice is also achieved with this technique. It offers an interesting complementary cooling mechanism for quantum simulations performed with quantum gases trapped in optical lattices. 
\end{abstract}
\maketitle

Atomic and molecular physics have been strongly impacted by cooling techniques \cite{BookCCTDGO,BookIF}. Laser cooling has triggered a boost of research activities to reach very low temperatures \cite{Chu,Cohen,Phillips} and therefore to improve the control on the external degrees of freedom of atoms, with many applications in metrology \cite{metrology1,metrology2}.  Laser cooling down to degeneracy has been demonstrated only recently \cite{grimm,vuletic}. Those cooling techniques are tailor-made for a given species. Lasers shall indeed address specific atomic lines. Furthermore, the temperature achieved with laser cooling is strongly dependent on the width of the excited state of the considered cycling transition \cite{Strontium}. In contrast, evaporative cooling is a much more generic technique. The filtering technique of the most energetic atoms can be easily transposed from one species to another, and the cooling occurs through the re-thermalization of the atomic cloud. Only the cooling rate depends on the species. Evaporative cooling has originally been envisioned for hydrogen atoms for which laser cooling is not expected to be efficient \cite{hess}, and successfully implemented on laser-cooled alkali atoms to reach quantum degeneracy in non-dissipative traps \cite{cornell,ketterle,kleppner}. 

In this article, we report on a generic cooling method inspired by evaporative cooling but adapted to atoms trapped in an optical lattice. It is indeed of upmost importance to find methods to decrease the temperature of atoms trapped in optical lattices since such systems are currently used to perform quantum simulations, and temperature may constitute an obstacle for some challenging experiments \cite{Bloch,Qin}. The advantage of optical lattices lies in their tunability: the geometry of the lattices can be easily modified, they can be made spin dependent and can be readily modulated in phase and amplitude. This latter possibility is at the origin of the so-called Floquet engineering which opens up interesting perspectives for generating effective hamiltonians and engineered artificial gauge fields \cite{RMPEckardt,Dalibard}.

The method put forward in this article exploits interband transitions excited by amplitude modulation for non-zero quasi-momentum \cite{SPECTRE}. The modulation frequency is scanned through all the values of the considered interband transition. The chosen excited band obeys a selection rule for its excitation from the ground state band which prohibits the transition of the wave function component with zero quasi-momentum. As a result atoms with the lowest quasi-momentum are not concerned by the excitation. In the following, we first present the experimental setup. We then show experimentally the effectiveness of selection rules. We also explain how parameters shall be tuned in order to benefit from the filtering operated by the frequency scan to remove the thermal wings of a partially condensed atomic cloud. The collisional dynamics which takes place inside the cloud after the removal of the most energetic atoms yields an increase of the fraction of condensed atoms, and a concomitant gain in phase space density. The method is eventually demonstrated on thermal atoms that are cooled down to degeneracy in this manner.

Our experiment starts with a rubidium-87 Bose Einstein Condensate (BEC) realized in a hybrid trap made of  a quadrupole magnetic trap and  two intersecting far-from-resonance optical dipole beams (wavelength 1064 nm, one horizontal and the other vertical)
\cite{PRL2016,PRA2018}. This setup can produce nearly pure condensates of about $10^5$ atoms. However, in this article, we have used uncondensed or partially condensed atomic cloud. We control the degree of degeneracy through the duration of the evaporative cooling ramp. Before loading the atoms into the optical lattice, we remove adiabatically the vertical beam of the dipole trap; we then ramp up in 30 ms a horizontal one-dimensional optical lattice realized by the interference of two counter-propagating laser beams with wavelength 1064 nm along the $x$ axis (yielding a lattice spacing of $d = 532$ nm) aligned with the dipole trap. The optical lattice has a depth, $V_0$, measured through the dimensionless parameter $s_0$: $V_0=s_0E_L$ with $E_L=h^2/(2md^2)$ \cite{footnote1}. The horizontal extra confinement provided by the dipole beam and the magnetic gradient of the hybrid trap is equivalent to a harmonic confinement of frequency 10 Hz.

Once the sample has been prepared, we apply an amplitude modulation of the lattice whose frequency is scanned in resonance with some excited bands of the static optical lattice. During the modulation which lasts a time duration $t_{\rm mod}$, the lattice potential experienced by the atoms reads
\begin{equation}
V(x,t)=-s_0 E_L (1+\varepsilon (t)) \cos^2\left( \frac{\pi x}{d} \right).
\end{equation}
The modulation of the lattice depth is $\varepsilon(t)=\varepsilon_0 \sin(2\pi\nu(t) t)$ with $\nu(t)=\nu_i + (\nu_f-\nu_i)t/t_{\rm mod}$. The amplitude $\varepsilon_0$ is slightly larger than 0.3 (see below). The instantaneous frequency $\nu(t)$ varies slowly in time with respect to the period of oscillation $1/ \nu(t)$: a typical modulation time is on the order of 10 ms while the typical modulation frequency is on the order of  10 to 30 kHz. 

The principle of the method is easily explained by assuming that the initial atomic cloud lies in the ground state band after the adiabatic loading into the optical lattice. It spreads on a continuum of Bloch states whose quasi-momenta are in the interval $k\in[-\Delta k/2; \Delta k/2]$. The cooling method that we propose in the following starts by promoting atoms with more energy than the mean energy to higher bands using resonant amplitude modulation \cite{SPECTRE}. (the coupling strength between bands $n$ and $p$ is proportional to the matrix element $V_{np}=|\langle n,k|cos^2(\pi x/d)|p,k\rangle|^2$ where $|n,k \rangle$ are the Bloch state). Those atoms are associated to the largest quasi-momenta of the interval of $k$. Once promoted they propagate along the lattice with a velocity dictated by the local slope $dE/dk$ of the excited band which is on the order of a few hundreds micrometers per second \cite{PRA2013}. Those excited atoms ultimately leave the sample. 
In this procedure, we filter out the most energetic atoms. To obtain a gain in phase space and/or recover equilibrium we rely on elastic collisions. 

\begin{figure}[t]
\centering
\includegraphics[width=\linewidth]{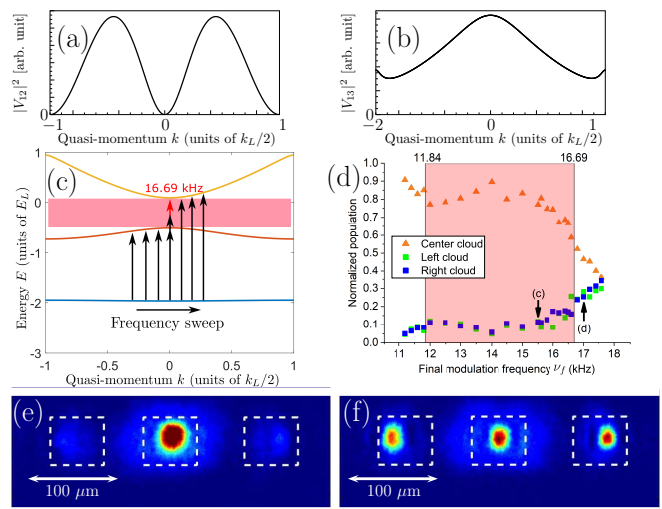}
\caption{(a) (resp.~(b)) Strength of the coupling, $|V_{12}|^2$ (resp.~$|V_{13}|^2$), between the ground state and the first (resp.~second) excited band as a function of the quasimomentum. (c) Sketch of the range of frequency spanned during the amplitude modulation superimposed to the band structure associated to an optical lattice of depth $s_0=2.7$. The gap between the first and second band is made visible with the pink area. (d) Summary of the data taken from the frequency $\nu_i=$11 kHz to an adjustable final frequency, $\nu_f$, ranging from 11.2 kHz to 17.6 kHz. The modulation time is $t_{\rm mod}=6$~ms and the modulation amplitude is fixed at $\varepsilon_0=0.38$. We have represented the number of atoms in the boxes delimited by the white dashed lines for the center and lateral peaks of the absorption images. Example of absorption images are provided for $\nu_f=15.5$ kHz (e) and $\nu_f=17$ kHz (f). Each image is averaged twice.}
\label{fig1}
\end{figure}

The robustness of our method originates from the use of selection rules, and therefore from the proper choice of bands over which the excitation is performed. Indeed, the selection rules for amplitude modulation at the center of the Brillouin zone ($k=0$) inhibit the promotion of atoms from the ground state band to the first (see Fig.~\ref{fig1}a) and third excited bands, and from the first excited band to the fourth one \cite{ZNA,SPECTRE}. However, this rule is only valid at the center of the Brillouin zone. We exploit this interesting feature to avoid the excitation of atoms from the condensate when we apply our protocol on a partially condensed cloud of atoms. Indeed, in this latter case, the condensate occupies a narrow range of $k$ about $k=0$ while the thermal part spans a larger $k$ interval on the two lowest bands (see below). Our method relies on a precise knowledge of the band structure and the lattice depth. In practice, we determined the lattice depths with a high accuracy using the shift method \cite{PRA2018}.

Before addressing cooling issues, we directly demonstrate the efficiency of the protection of the BEC against the modulation. For this purpose, the experiment is carried out on a cloud with a condensed fraction of 40 \% in a modulated lattice of depth $s=2.7$ and with a modulation amplitude, $\varepsilon_0=0.38$. The excitation from the ground state band to the first excited band occurs in the frequency range [9.88;11.84]  kHz and the lowest frequency to excite the second band is $\nu_{2 \rm min}=16.69$ kHz for our parameters. In practice, we scan the modulation frequency from  $\nu_i= 11$ kHz to an adjustable final $\nu_f$ (see Fig.~\ref{fig1}c) choosen in the frequency range [11.2;17.6] kHz. To get rid of the atoms that have been promoted on the excited bands, we accelerate them by ramping down the optical lattice in 25 ms. This trick is nothing else than the well-known band mapping technique \cite{bm1,bm2,bm3}. Atoms then evolve in the horizontal guide for $t_{\rm hold}=20$ ms, and an absorption image is taken after a 25 ms time-of-flight in the absence of any confinement. 
Two examples of such absorption images are depicted in Fig.~\ref{fig1}e and f. Figure \ref{fig1}e has been taken after the frequency sweep with final frequency $\nu_f=15.5$ kHz, a frequency below the minimum frequency, $\nu_{2 \rm min}$, required to excite the second band. We observe two lateral clouds that correspond to the excited atoms leaving the central cloud with a finite velocity. Interestingly, the frequency $\nu(t)$ has been resonant with the $k=0$ transition (11.84  kHz) from the ground state to the first excited band but we do not see any atoms from the condensate in the lateral clouds. This is to be contrasted with what happens when the final frequency $\nu_f$ is above $\nu_{2 \rm min}$. In this case, atoms at $k=0$ from the condensate are promoted to the second excited bands for which there is no selection rules that forbids the excitation of the $k=0$ component (see Figs.~\ref{fig1}a and b). This can be clearly seen on the absorption image of Fig.~\ref{fig1}f taken for $\nu_f=17$ kHz: the lateral clouds contain a Bose Einstein condensate. Furthermore, we observe that atom number in the central cloud decreases while the final frequency increases in the interval resonant with the first band (see Fig.~\ref{fig1}d). It stays nearly constant in the gap except at the close proximity below the frequency $\nu_{2 \rm min}$. This latter feature may be associated to the fluctuations of the lattice depth from shot to shot and/or the modification of the band structure that results from the resonant coupling between the ground and first excited bands. 

This first experiment confirms the effectiveness of the selection rules when amplitude modulation is applied. However, atoms that are promoted on the first excited band move towards the edges of the cloud quite slowly (on the order of 10 $\mu$m/s for our parameters) at a velocity governed by the slope of the dispersion relation. Furthermore those atoms cannot strictly speaking escape from the lattice envelope since they are propagating on a bounded band \cite{EPL2013,PRA2015} (see Fig.~\ref{fig1}c). We have therefore performed a second set of experiments where we promote atoms on higher excited bands that are unbounded. The choice of the excited band results from a trade-off. We want that the promoted atoms belong to an unbounded band but for a given amplitude of modulation $\varepsilon_0$, the probability of excitation decreases with the band index. We have finally chosen to carry out a quantitative study of the cooling mechanism by using resonant transitions from the ground state to the third excited bands with a bare lattice of depth $s_0=2.2$ and a modulation amplitude $\varepsilon_0=0.38$. 

We have chosen partially condensed atomic clouds with an initial temperature of 130$\pm$10 nK. This energy scale is to be compared with the energy width of the ground state band at a vanishing lattice depth equal to $h^2/8md^2$ i.e. 97 nK. We conclude that through the adiabatic loading of the lattice the uncondensed cloud will populate completely the ground state band and partially the first excited band. The modulation frequency varies from $\nu_i=25$ kHz to a final modulation frequency $\nu_f$ ranging from 27 to 37.5 kHz. Interestingly, this range of frequencies is resonant for all quasi-momenta of the ground band to the third excited band, \emph{and also} for promoting atoms of the first excited band to the fourth one but only in the quasi-momenta interval [-0.6 $k_L/2$,0.6 $k_L/2$].

\begin{figure}[t]
\centering
\includegraphics[width=\linewidth]{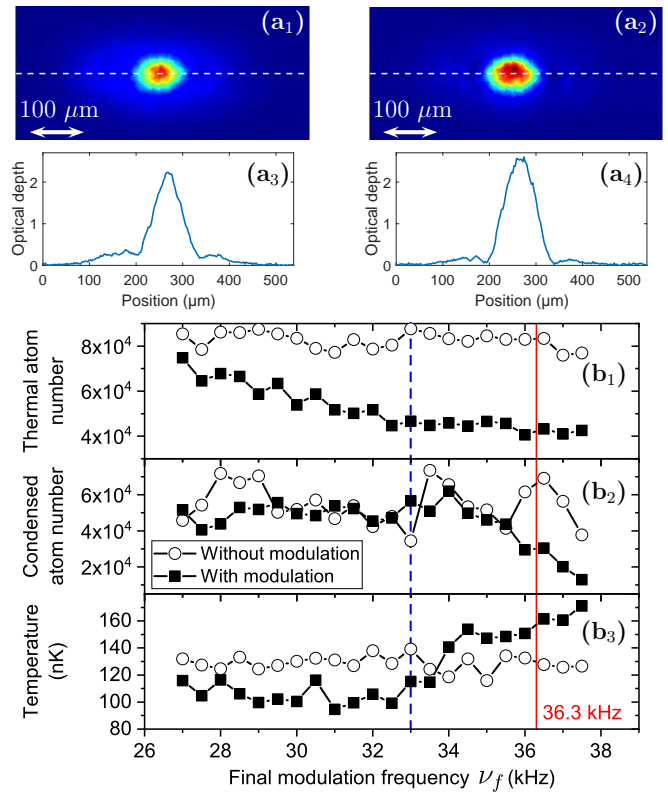}
\caption{Experimental results obtained for the amplitude modulation protocol with an optical lattice of depth $s_0=2.2$, a modulation amplitude $\varepsilon_0=0.38$ and a modulation time $t_{\rm mod}=10$ ms. The modulation frequency is scanned from $\nu_i=25$ kHz to an adjustable $\nu_f$ ranging from 27 to 37.5 kHz to excite atoms from the ground state band to the third excited band. ($a$) Example of results:  ($a_1$) Absorption image of a cloud of $3.4 \times 10^4$ condensed atoms and $8.8 \times 10^4$ thermal atoms at a temperature $T=115$ nK that has experimented the different stages of the protocol except for the modulation. (a$_2$): Absorption image obtained when applying the amplitude modulation protocol  for $\nu_f=33$ kHz.  (a$_3$) and (a$_4$): cut of both images along the white dashed line for both images. (b) Summary of all similar data: The number of thermal atoms ($b_1$) and condensed atoms ($b_2$) as well as the temperature ($b_3$) are depicted as a function of the final modulation frequency $\nu_f$. The vertical blue dashed line corresponds to $\nu_f=33$ kHz (see $a$) and the vertical red line identifies the frequency $\nu_{14}=36.3$ kHz associated to the transition from the ground state band to the third excited band at the center of the Brillouin zone for the bare lattice of depth $s_0$.}
\label{fig2}
\end{figure}

Figure \ref{fig2} summarizes our results on the atoms that are not excited by the modulation, analyzed after ramping down the lattice. We have represented as a function of $\nu_f$ the 
corresponding number of thermal atoms, the number of condensed atoms and their temperature. We clearly observe that the number of condensed atoms is roughly not affected by the amplitude modulation up to a final modulation frequency close to the resonant frequency $\nu_{14}=36.3$ kHz between the ground state band and the third excited band at the center of the Brillouin zone. By contrast the number of thermal atoms continuously decreases. This is an important result for realizing quantum simulations with optical lattices. Our method clearly demonstrates that the most energetic atoms in the lattice can be extracted from the lattice.

The temperature measured using the standard bimodal fit is clearly reduced for small final modulation frequency, a phenomenon that can be associated to the removal of the most energetic atoms as expected. However in an interval close to the resonant frequency the temperature starts to increase even above the initial temperature. The fact that this change of behavior is not exactly at $\nu_{14}$ can be explained by the large modulation amplitude which is about 40 \% of the bare lattice depth and which strongly modifies the lattice band structure. In the region for which the number of condensed atoms decreases ($\nu_f>36$ kHz) the modulation amplitude removes atoms with less energy than the mean energy and through the onset of thermalization the temperature is expected to increase. It means that our filtering protocol is accompanied by a thermalization process.
The interpretation of the temperature should nevertheless be taken with care since the system is probably not yet completely thermalized.

\begin{figure}[t]
\centering
\includegraphics[width=\linewidth]{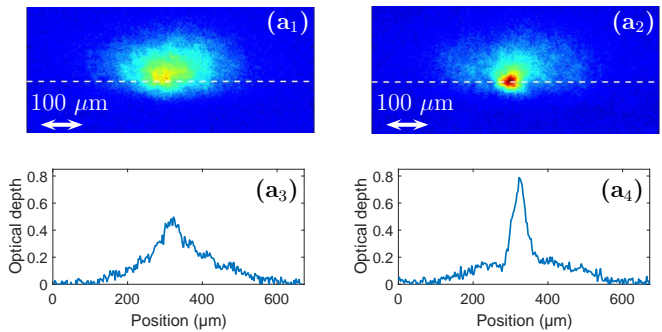}
\caption{(a$_1$): Absorption image of a cloud of $7 \times 10^4$ uncondensed atoms at a temperature $T=160$  nK that has experimented the different stages of the protocol except for the modulation. (a$_2$): Absorption image obtained when applying the amplitude modulation protocol (see text).  The cloud is now partially condensed ($2.8 \times 10^3$ condensed atoms and $5.6 \times 10^4$ thermal atoms). (a$_3$) and (a$_4$): cut of both images along the white dashed line.}
\label{fig3}
\end{figure}

As a result it shall be possible to use directly this protocol to increase the phase space density of the original cloud. 
For this purpose, we have done another set of experiments where we stop the evaporative cooling just above the condensation threshold in the hybrid trap. The vertical beam of the dipole trap is subsequently ramped down. At this stage, we have a cloud of $7 \times 10^4$ atoms at a temperature $T=160$  nK. For this experiment, we have chosen a lattice depth $s_0=1.8$, a modulation amplitude $\varepsilon_0=0.32$, and a modulation time $t_{\rm mod}=10$~ms. The frequency has been tuned from $\nu_i  = 25.7$ kHz to $\nu_f= 33.9$ kHz, driving the transition from the ground state band to the third excited band and from the first to the fourth excited bands \cite{footnoteF}. To analyze the data, we proceed as previously where atoms first propagate after the adiabatic ramping down of the lattice during $t_{\rm hold}=25$~ms into the horizontal guide of the dipole trap before the switching off of all confinements. The absorption images presented in Fig.~\ref{fig3} are taken after a 25 ms time-of-flight. We observe the emergence of a small condensate of $2.8 \times 10^3$ atoms surrounded by a thermal cloud of $5.6 \times 10^4$ atoms. 

This experiment confirms that our amplitude modulation protocol acts as evaporative cooling: a fraction of the most energetic atoms are filtered out by the modulation and the thermalization takes place. There is however a few important differences. Indeed the most energetic atoms selected by the modulation are promoted to excited bands whatever is their original position inside the lattice since the selection is made on the quasi-momentum. As a result, those atoms have to cross the sample and may collide with other atoms. Inside the lattice, the excited atoms have a velocity governed by the slope of the excited band over which they lie \cite{PRA2013}. If this velocity is below the superfluid velocity one expects the excited atoms to cross the BEC without deposing any energy \cite{superfluid}. The sound velocity is estimated to 3 mm/s and the velocity of the excited state is on the order of 100 $\mu$m/s in our case. However, in the band mapping procedure (adiabatic ramping down of the lattice), excited atoms are accelerated \cite{SPECTRE} above the sound velocity and dissipation may occur. Heating effects due to intra and interband transitions assisted by two particle processes are known to be problematic for 1D modulated lattice since collisions can deposit energy in the transverse degrees of freedom \cite{heat0,heat1,heat2,heat3,heat5,heat6,heat7}. This is to be contrasted with 3D lattices where this mechanism can be essentially inhibited. In our 1D modulated lattice, we have observed a clear cooling effect up to a lattice depth $s_0=4$, no more observable cooling effect at $s_0=7$ and even heating for a too long modulation time (80 ms). 

In conclusion, we have demonstrated an experimental technique that enables one to remove the most energetic atoms from an optical lattice. The method is generic and could be applied to cold fermions. Indeed, in this latter case, the possibility to empty the excited bands using a frequency scan of the amplitude modulation enables one to generate an insulator in 1D with a completely filled lowest band. This method can also be used to study out-of-equilibrium dynamics by removing selectively some classes of excited atoms. We have shown that this technique can be used to increase the phase space density of the trapped cloud thanks to the collisional dynamics. 
In this perspective, it is worth noticing that the cooling technique presented here is, to some extent, reminiscent of  the velocity-selective coherent population trapping (VSCPT) cooling scheme using a pair of Raman beams \cite{PRL88}. In the latter, the zero momentum velocity class is immune to the Raman light while other velocity classes undergo a Rabi oscillation between the coupled states that are contaminated by the excited state. As a result, atoms can decay by spontaneous emission. The corresponding randomization of the momentum can lead some atoms into the zero momentum class where they accumulate. In our protocol, most energetic atoms are removed. The subsequent thermalization increases the population by bosonic amplification in the phase space cell of the BEC, a state that is immune to amplitude modulation thanks to selection rules. Interestingly, one could cycle the protocol to get an even stronger cooling effect. It therefore offers a complementary cooling mechanism dedicated to quantum simulations performed in optical lattices. 

We thank Jean Dalibard for useful comments. This work was supported by Programme Investissements d'Avenir under the program ANR-11-IDEX-0002-02, reference ANR-10-LABX-0037-NEXT, and the research funding grant ANR-17-CE30-0024-01. M.~A. acknowledges support from the DGA (Direction G\'en\'erale de l'Armement).


\begin{thebibliography}{99}
\bibitem{footnote1}$E_L=4E_R$ where $E_R=h^2/(2 m \lambda^2)$ is the recoil energy associated to the optical lattice.
\bibitem{footnoteF}
Note that with a larger initial temperature a more optimal strategy would have consisted of promoting atoms from the populated excited bands towards even higher unbounded bands to avoid affecting two bands at the same time.
\end{thebibliography}
\end{document}